\documentclass[a4paper]{jpconf}
\usepackage{amsmath}
\usepackage{amssymb}
\usepackage{graphicx}
\usepackage{lineno}
\newcommand{\pIbi}        {{$10^{-2}$~cts/(keV$\cdot$kg$\cdot$yr)}}

\newcommand{\pIIbi  }     {{$10^{-3}$~cts/(keV$\cdot$kg$\cdot$yr)}}

\newcommand{\kgyr}        {{kg$\cdot$yr}}

\newcommand{\qbb}         {{$Q_{\beta\beta}$}}
\newcommand{\thalfzero}   {${T^{0\nu}_{1/2}}$}

\newcommand{\onbb}        {{$0\nu\beta\beta$}}
\newcommand{\nnbb}        {{$2\nu\beta\beta$}}

\newcommand{\gerda}       {\textsc{Gerda}}

\newcommand{\gess}        {{$^{76}$Ge}}
\newcommand{\geenr}       {{$^{\rm enr}$Ge}}          

\newcommand{\exposure}    {\mbox{$\cal E$}}

\begin{document}
\title{Searching for neutrinoless double beta decay with GERDA}

\author{
M.~Agostini$^o$,  
A.M.~Bakalyarov$^m$,  
M.~Balata$^a$,  
I.~Barabanov$^k$,  
L.~Baudis$^s$,  
C.~Bauer$^g$,  
E.~Bellotti$^{h,i}$,  
S.~Belogurov$^{l,k}$,  
A.~Bettini$^{p,q}$,  
L.~Bezrukov$^k$
% ...
}

\address{$^a$ INFN Laboratori Nazionali del Gran Sasso and Gran Sasso Science Institute, Assergi, Italy}
\address{$^b$ INFN Laboratori Nazionali del Sud, Catania, Italy}
\address{$^c$ Institute of Physics, Jagiellonian University, Cracow, Poland}
\address{$^d$ Institut f{\"u}r Kern- und Teilchenphysik, Technische Universit{\"a}t Dresden, Dresden, Germany}
\address{$^e$ Joint Institute for Nuclear Research, Dubna, Russia}
\address{$^f$ European Commission, JRC-Geel, Geel, Belgium}
\address{$^g$ Max-Planck-Institut f{\"u}r Kernphysik, Heidelberg, Germany}
\address{$^h$ Dipartimento di Fisica, Universit{\`a} Milano Bicocca, Milan, Italy}
\address{$^i$ INFN Milano Bicocca, Milan, Italy}
\address{$^j$ Dipartimento di Fisica, Universit{\`a} degli Studi di Milano e INFN Milano, Milan, Italy}
\address{$^k$ Institute for Nuclear Research of the Russian Academy of Sciences, Moscow, Russia}
\address{$^l$ Institute for Theoretical and Experimental Physics, Moscow, Russia}
\address{$^m$ National Research Centre ``Kurchatov Institute'', Moscow, Russia}
\address{$^n$ Max-Planck-Institut f{\"ur} Physik, Munich, Germany}
\address{$^o$ Physik Department and Excellence Cluster Universe, Technische  Universit{\"a}t M{\"u}nchen, Germany}
\address{$^p$ Dipartimento di Fisica e Astronomia dell'Universit{\`a} di Padova, Padua, Italy}
\address{$^q$ INFN  Padova, Padua, Italy}
\address{$^r$ Physikalisches Institut, Eberhard Karls Universit{\"a}t T{\"u}bingen, T{\"u}bingen, Germany}
\address{$^s$ Physik Institut der Universit{\"a}t Z{\"u}rich, Z{u}rich, Switzerland}

\ead{pandola@lns.infn.it}

\begin{abstract}
The GERmanium Detector Array (\gerda) experiment located at the INFN Gran Sasso Laboratory (Italy), is 
looking for the neutrinoless double beta decay of \gess, by using high-purity germanium detectors made from 
isotopically enriched material. The combination of the novel experimental design, the careful material 
selection for radio-purity and 
the active/passive shielding techniques result in a very low residual background at the $Q$-value 
of the decay, about \pIIbi.  This makes \gerda\ the first experiment in the field to be background-free 
for the complete design exposure of 100~\kgyr. 
A search for neutrinoless double beta decay was performed with a total exposure of 47.7~\kgyr: 23.2~\kgyr\ come 
from the second phase (Phase~II) of the experiment, in which the background is reduced by about a factor of  
ten with respect to the previous phase. 
The analysis presented in this paper includes 12.4~\kgyr\ of new Phase~II data.  No evidence for a possible 
signal is found: the lower limit for the half-life of \gess\ is $8.0 \cdot 10^{25}$~yr at 90\% CL.
The experimental median sensitivity is $5.8 \cdot 10^{25}$~yr. 
The experiment is currently taking data. As it is running in a background-free regime, its sensitivity 
grows linearly with exposure and it is expected to surpass $10^{26}$~yr within 2018.
\end{abstract}

\section{Introduction}
Neutrinoless double beta (\onbb) decay is a hypothetical nuclear transition in which a nucleus $(A,Z)$ 
decays as $(A,Z) \to (A,Z+2) + 2 e^-$. Neutrinoless double beta decay is forbidden by the Standard Model~(SM), 
as it violates by two units the conservation of the lepton number. A SM-allowed version of double beta decay 
also exists (\nnbb), in which the final state includes two anti-neutrinos. 
This rare SM process has been observed experimentally in about ten nuclei, with half-lives 
in the range of $10^{18}-10^{24}$~yr. 

\onbb\ decay is a far more interesting research topic than \nnbb, as it is a unique probe to physics beyond 
the SM. The observation of \onbb\ decay would prove in fact that neutrinos have a 
Majorana mass component~\cite{valle,lindner} and that lepton number is not conserved, thus shedding 
a new light on fundamental issues as the matter-antimatter asymmetry and the origin of 
neutrino masses~\cite{Davidson:2008bu,Mohapatra:2005wg,Pas:2015eia}. 
For this reason a wide experimental effort is placed worldwide in the search for 
neutrinoless double beta decay of several candidate nuclei, most notably \gess~\cite{nature,mjd}, 
$^{136}$Xe~\cite{exo,kz,exo2017} and $^{130}$Te~\cite{cuore0,snop}. 
The current best limits on the \onbb\ half-life 
lie in the range of $10^{24}-10^{26}$~yr, depending on the specific nucleus. It is 
clear that any experimental attempt to detect such long half-lives -- corresponding to decay rates smaller 
than 1~decay/(\kgyr) -- requires a very strong suppression of all background sources. This calls for the 
use of innovative low-background techniques (shielding, active vetoes, material selection for 
low-radioactivity) in a suitably-deep underground laboratory.

\section{The GERDA experiment}
The GERmanium Detector Array (\gerda) experiment~\cite{gerdatech} at the underground Laboratori Nazionali del Gran 
Sasso (LNGS) of INFN searches for the \onbb\ decay of \gess, using high-purity germanium (HPGe) detectors isotopically 
enriched to about 87\% in \gess. The experiment is located in the Hall 
A of LNGS, under a rock overburden of 3500~m w.e. 

The two electrons emitted in \onbb\ decay have a range of $\sim 1$~mm in 
germanium, so that in most cases their kinetic energy is entirely deposited within the detector itself. 
The experimental signature is hence a mono-energetic energy release at the $Q$-value of the decay, which 
is  $Q_{\beta\beta}=2039$~keV for \gess. A key feature of HPGe detectors is the excellent energy resolution, 
better than 4~keV full width at half maximum (FWHM) at \qbb. The very good energy resolution allows 
for a narrow search region for the \onbb\ signal, thus minimizing the background, and 
for the convincing identification of a narrow line at the expected energy, in case of a discovery. 

\begin{figure}[tb]
\begin{center}
\includegraphics[height=0.27\textheight]{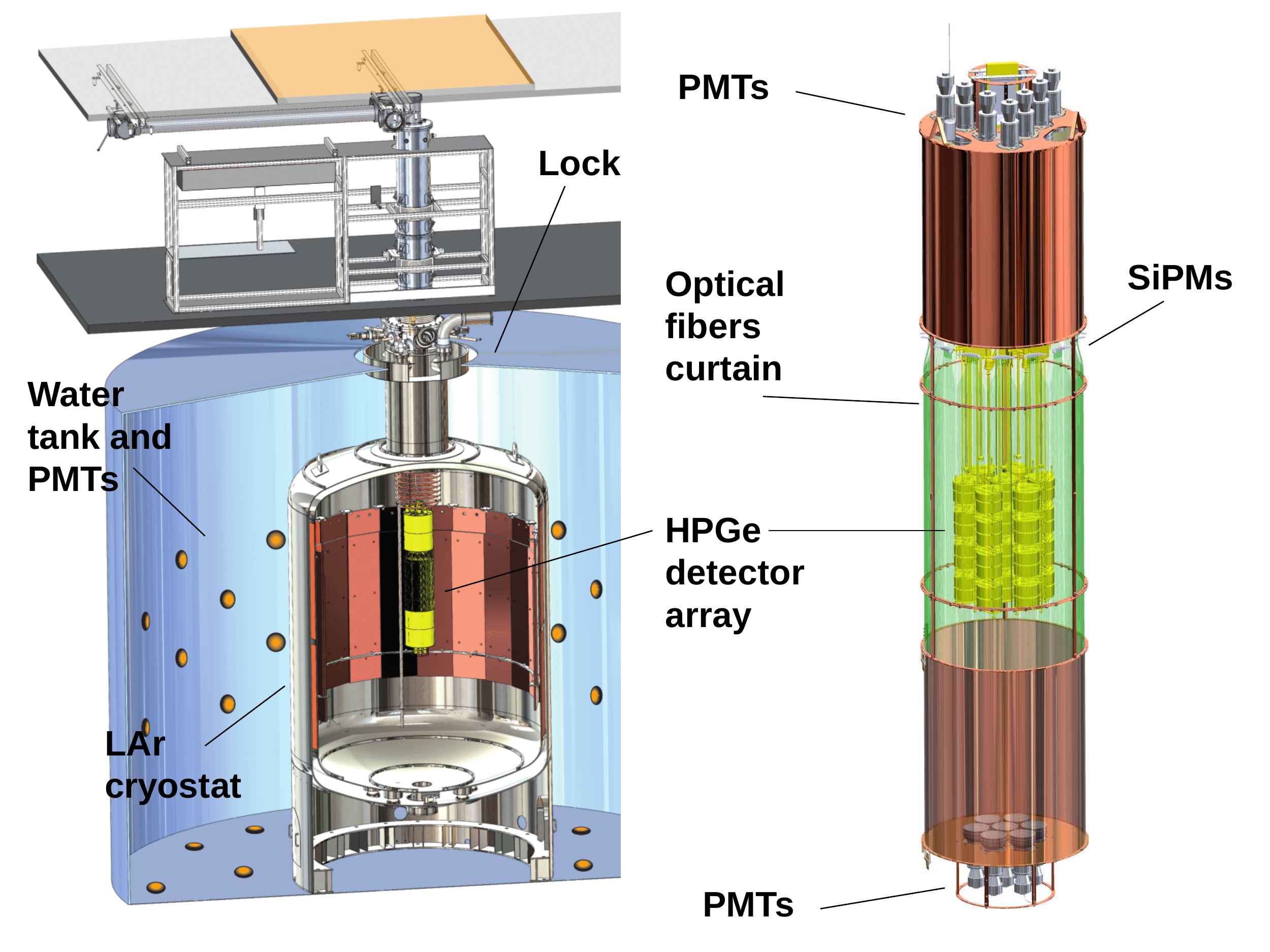}
\caption{\label{fig:array}  Schematic layout of the \gerda\ setup. The right-hand side of the figure 
displays a zoom of the inner part of the setup.}
\end{center}
\end{figure}

The key design feature of \gerda\ is the operation of bare HPGe detectors directly immersed in 
liquid argon (LAr). Liquid argon, contained in a 64~m$^3$ cryostat, acts as 
cooling medium and as a shielding against the external radiation. The cryostat is 
enclosed in a stainless steel water tank containing 590~m$^3$ of ultra-pure water, serving as additional 
shielding and effective neutron moderator. The water tank is equipped with photomultipliers, such that the 
water volume is operated as a Cerenkov detector to veto the residual cosmic ray muons in the 
underground laboratory. The HPGe detectors are arranged as a 
closely-packed array of seven strings, that are deployed in the LAr cryostat through a neck on the top, as shown 
in Fig.~\ref{fig:array}. Two sets of detectors enriched in \gess\ (\geenr) are currently operated in \gerda: 
refurbished p-type coaxial detectors, that were previously operated by the 
Heidelberg-Moscow~\cite{KlapdorKleingrothaus:2000sn} 
and IGEX~\cite{igex} experiments; and new Broad Energy Germanium (BEGe) p-type detectors, manufactured by 
Canberra. The BEGe detectors are designed to provide much superior pulse shape discrimination 
performance with respect to the old-style coaxial detectors~\cite{dusan,psd}. 
The charge signals from the HPGe detectors are sampled 
for a total window of 160~$\mu$s and stored on disk. Data are analyzed offline using the 
\textsc{Gelatio} software package~\cite{gelatio} and the procedures described in ~\cite{prl,acat}. 

As LAr is a very effective scintillator, the LAr volume surrounding the detector array 
is equipped with photo-detectors and operated as an active veto. In particular, the HPGe strings are enclosed 
by a curtain of scintillating fibers coated with a wavelength shifter, in order to convert the wavelength 
from the original 128~nm of the argon scintillation to the visible range. The light collected by the fibers 
is detected by 90 SiPM, arranged in 15 read-out channels. The LAr light detection system is complemented by a set 
of 16 low-radioactivity photomultipliers, placed above and below the HPGe strings. 

The experiment had been designed to proceed through two phases. The first phase (Phase~I), from 
November 2011 to May 2013, accumulated an exposure of 21.6~\kgyr\ with a background of about 
\pIbi\ at \qbb~\cite{prl}.
The LAr volume was not instrumented with photo-sensors and it was 
hence used as a pure passive shield. The second phase (Phase~II) is ongoing since December 2015: 
it features the full instrumentation of the LAr volume and the doubling of the mass of \geenr\ 
detectors, by the deployment of new BEGe detectors. In total, 37 \geenr\ detectors 
are available in Phase~II, 30 BEGe (20.0~kg) and 7 coaxial (15.6~kg). The goal of Phase~II is to accumulate 100~\kgyr\ of 
exposure with a background of \pIIbi, i.e. an order of magnitude smaller with respect to Phase~I.
As in Phase~I, the data taking  
follows a blind analysis strategy: all events in the Ge detectors with reconstructed energy within 
\qbb$\pm$25~keV, i.e. close to the expected signal region, are stored on disk but not made available 
for analysis. The blinding is released approximately every year, after all 
analysis cuts are tested and finalized.  

The initial Phase~II results were released in June 2016 with 
10.8~\kgyr\ total exposure, indicating that the ambitious background goal has indeed been met~\cite{nature}. 
The expected background in the signal search region 
\qbb$\pm$0.5~FWHM is $\lesssim 1$~count for the full design exposure, thus making \gerda\ to be 
the first ``background-free'' experiment in the field. The sensitivity is expected to grow linearly with 
the exposure \exposure\ for the entire data taking period, instead of the $\sqrt{\cal E}$ regime 
of background-dominated experiments. 
%

\section{Data taking and selection}
The Phase~II data taking is ongoing since December 25$^{th}$, 2015. Data collected up to April 15$^{th}$, 
2017 (477.2~calendar days) have been fully validated and analyzed, totaling an \geenr\ exposure of 
34.4~\kgyr\ (18.2~\kgyr\ from BEGe detectors and 16.2~\kgyr\ from coaxial detectors). 
The exposure is hence increased by a factor of three with respect to the previous 
Phase~II release~\cite{nature}. 
%
%

\begin{figure}[tb]
(a)
\includegraphics[height=0.23\textheight]{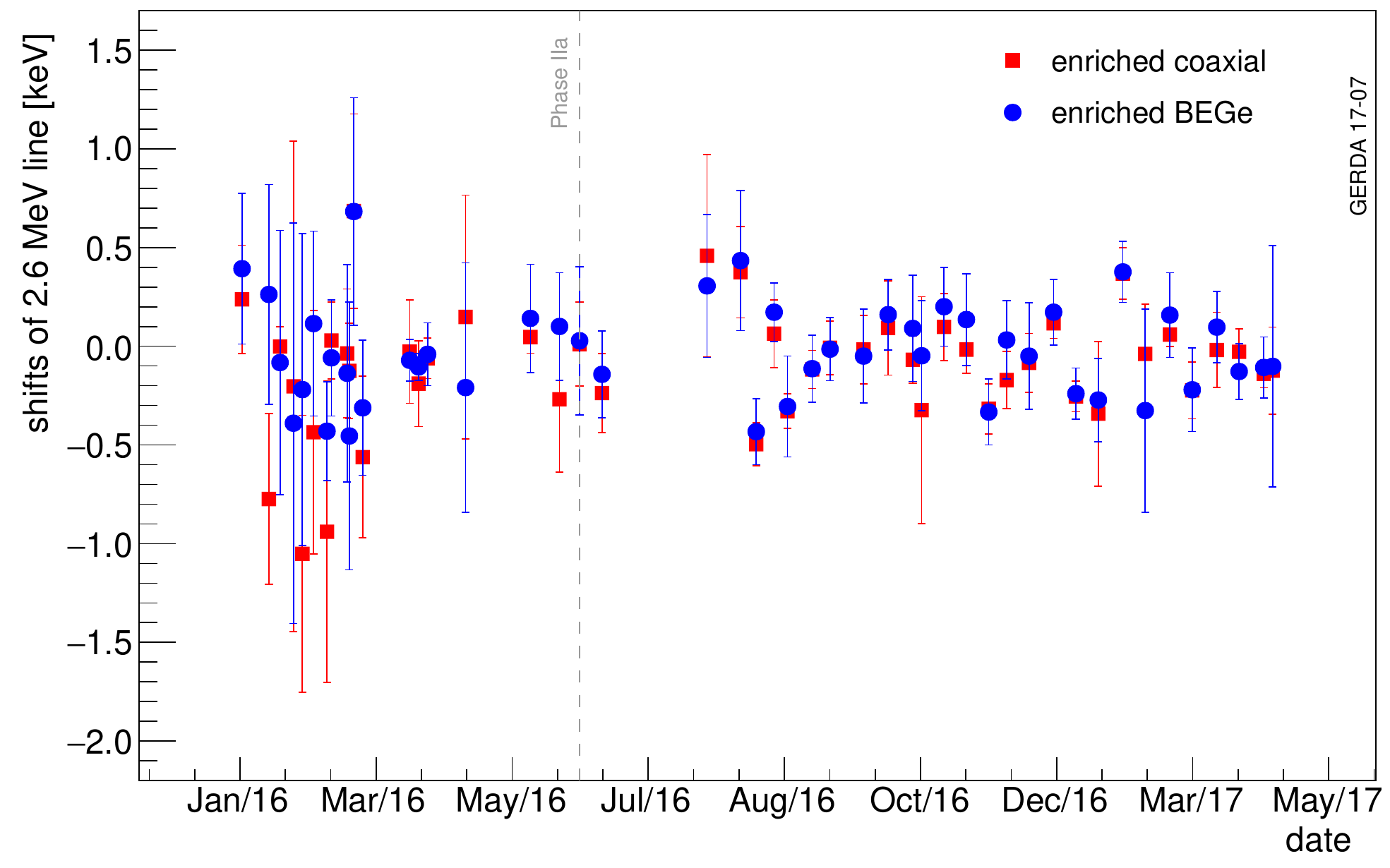}
(b)
\includegraphics[height=0.23\textheight]{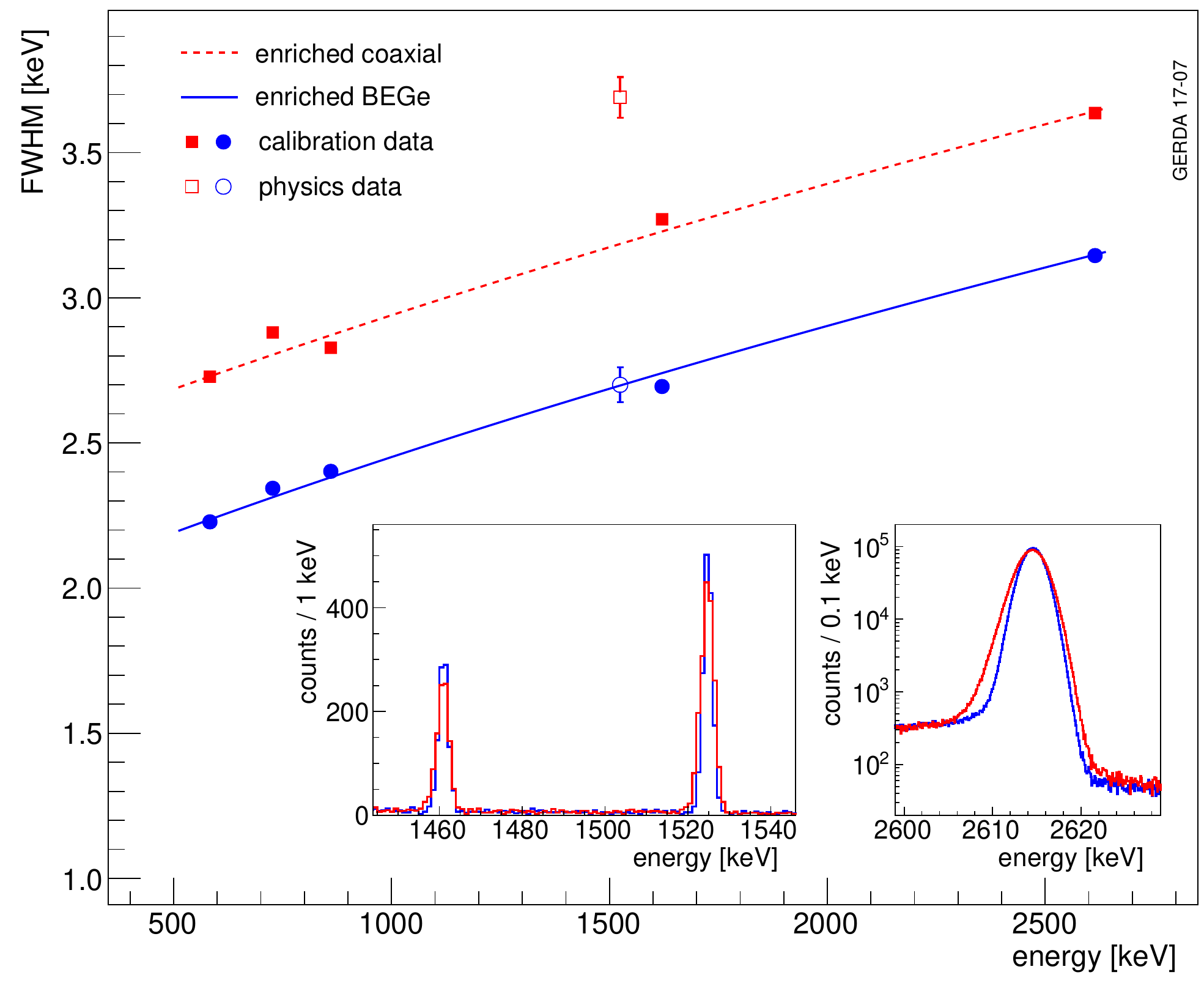}
\caption{\label{fig:cal}  (a) Average shift of the 2615~keV $\gamma$-ray line from the $^{228}$Th between consecutive 
calibrations. The error bars represent the standard deviation of the shifts of the individual detectors. 
(b) Average energy resolution for $\gamma$ lines observed in calibration data and in physics data, 
for the BEGe and coaxial detectors. The inset displays a zoom of the $\gamma$ lines at 1460~keV ($^{40}$K) 
and 1525~keV ($^{42}$K) in physics data, and a zoom of the 2615~keV line from calibration data.}
\end{figure}

The Ge event energy is reconstructed offline using the Zero Area Cusp (ZAC) algorithm~\cite{zac}, which is 
optimized for each detector and each calibration individually. The energy scale is set 
by taking weekly calibration runs with $^{228}$Th sources. 
Given the excellent energy resolution of HPGe detectors, the energy scale at \qbb\ must be stable within   
$\lesssim 2$~keV throughout the data taking. The stability of the system is monitored weekly, by checking the 
shifts of the $\gamma$-lines between two consecutive calibrations, and continuously, by injecting 
charge pulses of fixed amplitude (test pulses) every 20~s. Periods in which the energy scale exhibits
jumps or drifts exceeding 2~keV are discarded from the analysis. The remarkable stability of the 
2615~keV $\gamma$-line from the $^{228}$Th source between two consecutive calibrations is displayed in 
Fig.~\ref{fig:cal}a.
The energy resolution at \qbb\ is also evaluated by using the calibration data. Resolution vs. energy 
is parametrized as $\sigma(E) = \sqrt{a + bE}$; the free parameters $a$ and $b$ are evaluated, 
for coaxial and BEGe detectors separately, by a fit of the 
$^{228}$Th $\gamma$-lines (see Fig.~\ref{fig:cal}b). Nevertheless, it is observed for the coaxial detectors 
that the energy resolution achieved in the physics data for the $^{40}$K and $^{42}$K lines (1460 and 1525~keV, 
respectively) is slightly worse than the expectation based on calibrations. This could be due to 
long-period gain fluctuations and instabilities taking place during the multi-week physics runs.
The effect 
is accounted for by including a further summation term in the formula above, such to match the 
observed resolution of the $^{40}$K and $^{42}$K peaks in the physics data. No correction is required 
for the BEGe detectors, as shown in Fig.~\ref{fig:cal}b. The average resolution at \qbb\ is 3.90(7)~keV and 2.93(6)~keV 
FWHM for the \geenr\ coaxial and BEGe detectors, respectively.

%
Unphysical events, e.g. due to discharges, are effectively removed by means of a set of dedicated quality 
cuts based on features of the digitized traces, as baseline flatness, polarity and time structure. The 
loss of genuine \onbb\ signals due to mis-classification by the quality cuts is estimated to be smaller than 0.1\%, based  
on the acceptance of $\gamma$-lines in calibration data, test pulse events and simulated template signals.  
%
The peculiar topology of \onbb\ decay  and the signature 
which is used for the search (i.e. full energy deposition at \qbb) allow to define efficient analysis cuts, 
featuring high selection efficiency and high ability to reject the $\alpha$ and $\gamma$ background. 
The target events of \gerda\ are those in which 
the whole energy \qbb\ is released within one HPGe detector, which happens for about 83\% of decays taking place 
in the detector itself. The probability above accounts for the presence of an inactive volume inside the detectors 
and for the possibility that energy partially escapes, e.g. via bremsstrahlung or x-rays. 
Events in which more than one Ge detector has an energy 
deposit, or in which an energy deposit in liquid argon is detected within  5~$\mu$s from the Ge 
signal, are rejected as background events. The LAr-veto inefficiency is ($2.3\pm0.1$)\%, due to random 
coincidences which falsely produce the rejection of a \onbb\ decay.
Furthermore, Ge events in coincidence with a muon veto signal within a 10~$\mu$s window are also discarded. 

Genuine \onbb\ decays produce a very localized energy deposit, due to the short range of electrons in Ge, 
and are predominantly single-site events (SSEs).
The most important 
background sources are either $\gamma$-rays, which mostly interact by multiple Compton scattering separated 
by $\sim 1$~cm in Ge (multi-site event, MSE), or external $\alpha$/$\beta$-rays, which deposit their energy 
on the detector surface. Given this difference, the time profile of the Ge current signal can be used to discriminate 
single-site events against multi-site or surface background events. The selection cuts 
are set for each detector individually; they are defined by using the double escape peak (DEP) of the 2615~keV line of 
the $^{228}$Th calibrations as a proxy for signal events. 

The electric field configuration of BEGe detectors allows for a very effective mono-parametric 
pulse shape discrimination (PSD), based on the ratio $A/E$ between the maximum amplitude $A$ of the 
current signal and the total energy $E$~\cite{dusan,psd,matteosim}. The $A/E$ parameter is corrected 
to account for time and energy dependence and normalized as described in Ref.~\cite{vicithesis}. 
The cut on $A/E$  allows to reject $> 90\%$ of ($\gamma$-like) MSEs 
and basically all $\alpha$-like surface events. The \onbb\ selection efficiency 
is $(87 \pm 2)$\%. 
Pulse shape discrimination is less effective for the coaxial detectors and cannot be performed 
with a single parameter~\cite{psd}. Two artificial neural network (ANN) algorithms are instead used, to 
classify SSE vs. MSE and internal vs. $\alpha$ surface events, respectively. 
The first network 
is trained with calibration events from the DEP of $^{208}$Tl at 1592~keV (SSE sample) and from the 
nearby $\gamma$-line at 1620~keV from $^{212}$Bi (MSE sample), respectively. The second network is trained 
using \nnbb\ events (internal sample) and $\alpha$ events (surface sample) from physics data.
The selection efficiency of the ANNs is evaluated by applying the selection to a sample of \nnbb\ events 
and to traces simulated by a dedicated Monte Carlo code~\cite{akthesis}. The 
combined selection efficiency for \onbb\ decays is $(79 \pm 5)$\%. 

%
%
\section{Statistical analysis and results}
Data from the BEGe detectors taken between June 1, 2016 and April 15, 2017 have been disclosed 
in June 2017, totaling 12.4~\kgyr\ of additional exposure with respect to Ref.~\cite{nature}. Only two extra events are 
found which pass all selection cuts; both of them are more than 15~keV away from \qbb\ 
(namely $> 10 \sigma$) and cannot hence be attributed to \onbb\ decay. Data from the coaxial 
detectors (11.2~\kgyr) were not unblinded because it was recently realized that a fraction of 
$\alpha$-induced events, that are characterized by a very short rise time 
are not correctly identified by the ANN PSD and leak as candidate \onbb\ events. As these events are 
clearly background, data are to be unblinded only when the PSD is upgraded and the 
new cut is frozen. 
 

The exposure available for analysis is 23.5~\kgyr\ and 23.2~\kgyr\ for Phase~I and Phase~II, respectively, 
totaling $(236.9 \pm 4.6)$~mol$\cdot$yr of \gess\ in the active volume of the detectors.  
Data are grouped into six data sets, according to the detector type 
(and energy resolution) and to the background at \qbb, as listed in Table~\ref{tab:datasets}. The 
statistical analysis is described in the Methods Section of Ref.~\cite{nature}: it is performed as a simultaneous 
unbinned maximum likelihood fit of the energy spectrum of the 
six data sets $D_i$ in the range between 1930 and 2190 keV. The fit range excludes the intervals 
$(2104\pm5)$~keV and $(2119\pm5)$~keV, where known background peaks are expected, such that it spans 
$\Delta E = 240$~keV in total. For a given \onbb\ signal strength $S$=1/\thalfzero\, the expected number of 
identified signal events in a data set is given by:
\begin{math}
\mu_i^S = \frac{\ln 2 \mathcal{N}_A}{m_a} \epsilon_i \mathcal{E}_i S 
\end{math}
where $\mathcal{N}_A$ is Avogadro's number, $m_a$ the molar mass, $\epsilon_i$ the total efficiency and 
\exposure$_i$\ the exposure (i.e. detector mass $\times$ time). The efficiency $\epsilon_i$ 
accounts globally for: the enrichment fraction of \gess\ in \geenr; the probability that the entire decay energy 
\qbb\ is released in the active fraction of the Ge detector; the efficiency of the offline reconstruction and of 
all cuts. Similarly, the expected number of background events surviving the selection cuts is calculated as
\begin{math}
\mu_i^B =  \mathcal{E}_i B_i \Delta E 
\end{math}
where $B_i$ is the background index. 
Each data set spectrum is fitted with a function made by a Gaussian distribution at \qbb\ for the 
signal and a flat distribution for the background. 
The common signal strength $S$ and all backgrounds $B_i$ are bound to positive values. 

\begin{table}
\caption{\label{tab:datasets} Summary of the Phase~I and Phase~II analysis data sets (exposure, energy resolution, 
total efficiency and background index). Background is evaluated in the range between 1930 and 2190 keV, 
excluding the intervals $(2104\pm5)$~keV and $(2119\pm5)$~keV, where known background peaks are expected, 
and the signal interval $(Q_{\beta\beta} \pm 5)$~keV.}
\begin{center}
\begin{tabular}{lcccc}
\br
Data set & Exposure & Resolution at \qbb & Efficiency & Background Index \\
 & (\kgyr) & (keV) FWHM & $\epsilon$ & \pIIbi \\
\mr
Phase~I golden~\cite{prl} & 17.9 & 4.3(1) & 0.57(3) & $11\pm2$ \\
Phase~I silver~\cite{prl} & 1.3 & 4.3(1) & 0.57(3) & $30\pm10$ \\
Phase~I BEGe~\cite{prl} & 2.4 & 2.7(2) & 0.66(2) & $5^{+4}_{-3}$ \\
Phase~I extra~\cite{nature} & 1.9 & 4.2(2) & 0.58(4) & $5^{+4}_{-3}$ \\
\quad Total Phase I & 23.5 & & & \\
\hline
Phase~II coaxial~\cite{nature} & 5.0 & 4.0(2) & 0.53(5) & $3.5^{+2.1}_{-1.5}$ \\
Phase~II BEGe & 18.2 & 2.93(6) & 0.60(2) & $1.0^{+0.6}_{-0.4}$ \\
\quad Total Phase~II & 23.2 & & & \\
\hline
\quad Total Phase~I and II & 46.7 & & & \\
\br
\end{tabular}
\end{center}
\end{table}
The same likelihood function is used to 
perform the analysis in a frequentist and in a Bayesian 
framework. The frequentist analysis employs a test statistics $t_S = -2 \ln \lambda(S)$ 
based on the profile likelihood $\lambda(S)$~\cite{cowan}. Due to the low-statistics regime of \gerda, Wilks's 
theorem is not valid, typically yielding under-coverage. The confidence intervals are derived by using a  
toy Monte Carlo approach, which ensures the correct coverage by construction. 
For each assumption of the signal strength $S$ the full distribution $f(t_S)$ of the test statistics is built and the 
$p$-value for the observed data is calculated. The analysis of the \gerda\ data returns a best-fit $S=0$ (i.e. no signal). 
The limit on the half-life of \gess\ is \thalfzero $> 8.0 \cdot 10^{25}$~yr (90\% CL). 
The median sensitivity for the 90\% CL lower limit of \thalfzero\ is $5.8 \cdot 10^{25}$~yr. It is calculated by 
producing Monte Carlo realizations of \gerda\ with different signal strengths and evaluating the test-statistics 
for the null hypothesis. The probability to produce a limit 
stronger than the actual one within an ensemble of Monte Carlo \gerda\ realizations is about 30\%.

A statistical analysis within the Bayesian framework is also available. The posterior probability of the signal strength 
$P(S)$ is calculated according to Bayes's theorem, after 
marginalization of all other nuisance parameters. 
A flat prior between 0 and $10^{-24}$~1/yr is taken for the signal strength. 
The 90\% credible interval obtained from the analysis is \thalfzero $ > 5.1 \cdot 10^{25}$~yr;  
it is calculated as the minimum interval of the 
posterior probability $P(S)$ which contains 90\% of the probability.
The median sensitivity is
\thalfzero $ > 4.5 \cdot 10^{25}$~yr;  
as for the frequentist analysis, it is evaluated by repeating the full analysis procedure for a large set of Monte Carlo 
realizations of the experiment. 
%
%
\begin{figure}[tbp]
\begin{center}
\includegraphics[height=0.32\textheight]{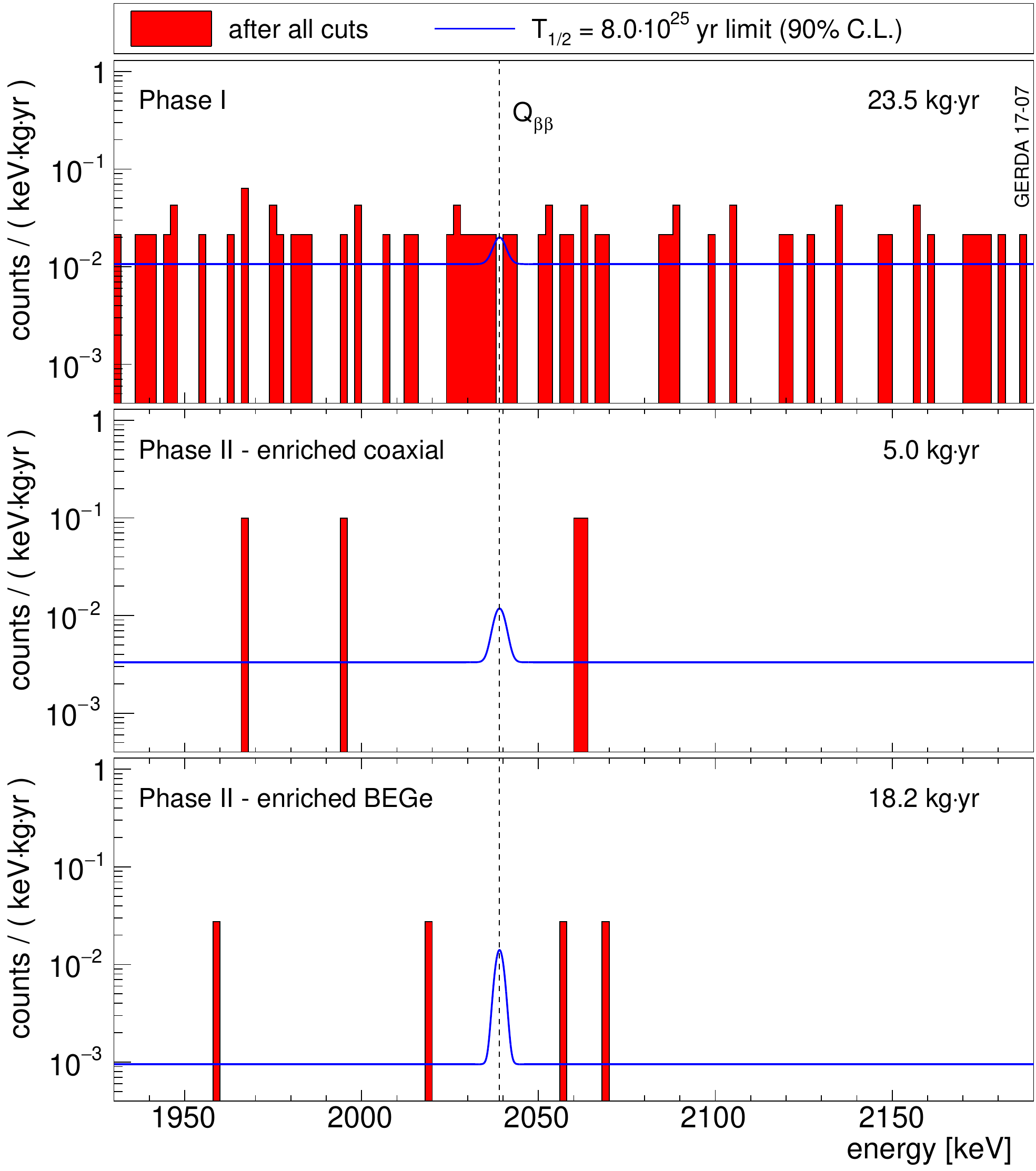}
\caption{\label{fig:roizoom}  Energy spectra around \qbb\ for Phase~I, Phase~II coaxial detectors 
and Phase~II BEGe detectors after all cuts. The binning is 2~keV. The blue lines show
the hypothetical \onbb\ signal for \thalfzero$=8.0\cdot 10^{25}$~yr, sitting on the 
constant background.}
\end{center}
\end{figure}



\section{Discussion and perspectives}
\gerda\ is currently taking data, featuring an extremely low background at \qbb\ of \pIIbi. 
The background is low enough to make 
the experiment background-free for the entire design exposure of 100 \kgyr\ and to make the sensitivity grow linearly 
with exposure.
In spite of the much smaller exposure, 
the \gerda\ physics performance is comparable with the leading $^{136}$Xe experiments, thanks
to the superior background level and energy resolution.
The sensitivity on \thalfzero\ reported here for \gess\ ($5.8 \cdot 10^{25}$~yr), 
is slightly better than what is obtained by KamLAND-Zen for $^{136}$Xe, $5.6 \cdot 10^{25}$~yr~\cite{kz}. Taking the unblinded 
data into account, \gerda\ has currently the 
record sensitivity on \thalfzero. The Phase~II background, normalized to the energy resolution and to the total efficiency 
(i.e. $B \cdot$FWHM/$\epsilon$), is 5 (20) counts/(ton$\cdot$yr) for BEGe (coaxial) detectors, namely a factor of five 
- or more - better than any competing non-\gess\ experiment.

\gerda\ is currently accumulating data. Taking into account the 
background level achieved so far and 11.2~\kgyr\ of still-to-be-unblinded coaxial data, 
the median sensitivity will break the wall of  $10^{26}$~yr within 2018. 
The sensitivity expected for the 
full 100~\kgyr\ exposure of \gerda\ is about $1.4 \cdot 10^{26}$~yr. Furthermore, the excellent energy resolution 
makes \gerda\ very well suited for a possible discovery of \onbb\ decay: with \exposure=100~\kgyr, a $3\sigma$ evidence of 
\onbb\ decay could be claimed with 50\%-chance for \thalfzero\ up to $\sim 8 \cdot 10^{25}$~yr. A discussion of 
discovery potentials for present and future \onbb\ experiments can be found in Ref.~\cite{matteodiscovery}.

After the end of \gerda\ 
Phase~II in 2019, the newly formed \textsc{Legend} Collaboration plans to operate up to 200~kg of 
\geenr\ detectors in the existing \gerda\ infrastructure at Gran Sasso~\cite{legend}. 
In order to maintain the background-free regime up to 1000~\kgyr\ exposure, a background 
reduction by a factor of 5-10 with respect to \gerda\ is aimed for. The second stage of 
\textsc{Legend} would be a 1-ton \gess\ background-free experiment: 
such an ultimate experiment would reach a sensitivity of $10^{28}$~yr, 
and hence fully cover the inverted hierarchy region, in ten years of data taking. 


\section*{References}
\bibliography{biblio}{}
\bibliographystyle{iopart-num}

\end{document}